\documentclass{article}

\usepackage{arxiv}

\usepackage[utf8]{inputenc} 
\usepackage[T1]{fontenc}    
\usepackage{hyperref}       
\usepackage{url}            
\usepackage{booktabs}       
\usepackage{amsfonts}       
\usepackage{nicefrac}       
\usepackage{microtype}      
\usepackage{lipsum}		
\usepackage{graphicx}
\usepackage{natbib}
\usepackage{doi}
\usepackage{comment}

\title{``Nudes? Shouldn't I charge for these?'':\\Motivations of New Sexual Content Creators on OnlyFans}

\author{
\textbf{Vaughn Hamilton}\\
\texttt{vhamilto@mpi-sws.org} \\
Max Planck Institute for Software Systems
\and
\textbf{Ananta Soneji}\\
\texttt{asoneji@asu.edu}\\
Arizona State University
\AND
\textbf{Allison McDonald} \\
\texttt{amcdon@bu.edu} \\
Boston University \\ 
\and
\textbf{Elissa M. Redmiles} \\
\texttt{eredmiles@mpi-sws.org} \\
Max Planck Institute for Software Systems
}

\hypersetup{
pdftitle={``Nudes? Shouldn't I charge for these?'':\\Motivations of New Sexual Content Creators on OnlyFans},
pdfauthor={Vaughn},
pdfkeywords={OnlyFans, gig work, digital sex work, platform labour, pandemic, digital labor, informal labor, sexual expression, computer-supported cooperative work},
}

\begin{document}
\maketitle

    \begin{abstract}
With over 1.5 million content creators, OnlyFans is one of the fastest growing subscription-based social media platforms. The platform is primarily associated with sexual content. Thus, OnlyFans creators are uniquely positioned at the intersection of professional social media content creation and sex work. While the experiences and motivations of experienced sex workers to adopt OnlyFans have been studied, in this work we seek to understand the motivations of creators who had not previously done sex work. Through a qualitative interview study of 22 U.S.-based OnlyFans creators, we find that beyond the typical motivations for pursuing gig work (e.g., flexibility, autonomy), our participants were motivated by three key factors: (1) societal visibility and mainstream acceptance of OnlyFans; (2) platform design and affordances such as boundary setting with clients, privacy from the public, and content archives; and (3) the pandemic, as OnlyFans provided an enormous opportunity to overcome lockdown-related issues.
    \end{abstract}

\section{Introduction}
Social media platforms are innovating their business models to improve engagement of both viewers and content creators. To engage creators in developing high quality content that will drive viewership, platforms are increasingly facilitating methods for
creators to be paid for their work through a variety of different monetization models~\cite{kopf2020rewarding}. Some platforms, such as YouTube and TikTok, have created programs to pay creators in proportion to the views on their content, while others like Twitch, Patreon, and OnlyFans leverage a subscriber model in which fans pay creators directly through subscriptions, tips and gifts.

The experiences of creators on these platforms have a deep impact on the success of the platform. Thus, to inform platform design and improve labor rights for a growing sector of gig workers---professional content creators---prior work has focused on the motivations of professional YouTube, Twitch, and Patreon creators, as well as the moderation and monetization schemes, harassment experiences, and subscriber motivations and relationships on these platforms, among other topics~\cite{wohn2019understanding,sheng2020virtual,johnson2019s,taylor2018watch,seering2018social,torhonen2019fame,regner2021crowdfunding}.

OnlyFans is among the newest subscription-based social media platforms. Launched in 2016, OnlyFans is an online subscription platform which creators use to earn money for their content.
Users, also known as fans, subscribe to the content of these creators.
OnlyFans creators can receive funds from their fans via monthly subscription fees, tips, paid private messages, and the pay-per-view feature~\cite{bonifacio2020digital, bernstein2019onlyfans, hall2018onlyfans}.

OnlyFans is unique from its predecessors in that, while the platform advertises that it hosts a wide variety of content, it is primarily associated with adult content (i.e., sexual photos and videos that are protected by a paywall), rather than game streaming or artistic creation~\cite{van2021competing}. 
In contrast to other social media platforms (e.g., Twitch, YouTube, Patreon, Facebook), OnlyFans' terms of service are relaxed for adult content creation~\cite{bonifacio2020digital, hall2018onlyfans}. 
Thus, OnlyFans is uniquely positioned between the spaces of digital sex work and subscription-based social media. Further, OnlyFans experienced rapid growth during the COVID-19 pandemic: revenue grew by over 500\%~\cite{OnlyFans29:online} and is now estimated to equal that of Twitch, despite having fewer subscribers.\footnote{OnlyFans has an estimated $2$ million subscribers vs. Twitch's $7$ million~\cite{OnlyFans57:online}. Estimates by Statista (\url{https://www.statista.com/}) put both platforms' revenue around $2.5$ billion for 2022.}

Economists, psychologists, social scientists and researchers of computer-supported cooperative work have long sought to understand individual worker motivations for pursuing particular forms of labor~\cite{kanfer2017motivation}. Most relevant to our particular study, prior work has examined what motivates professional creators on other social media platforms, people's choice to pursue various forms of gig work during the COVID-19 pandemic, as well as how sex workers adopt and navigate new commercial sexual platforms including OnlyFans~\cite{dunn2020motivation, cardoso2021bodies, hamilton2022risk, vallas2020platforms, matikainen2015motivations, umar2021impact, cano2021flexibility, katta2020dis,apouey2020gig,dunn2020motivation}. However, less is known about what draws people who have never previously worked in the sex industry to begin creating content on platforms like OnlyFans. Further, OnlyFans experienced an abrupt spike in popularity and visibility: after Beyoncé mentioned the site in a remix in April, OnlyFans reported ``a $15$ percent spike in traffic'' in the subsequent $24$ hours.\footnote{https://www.thedailybeast.com/adult-site-onlyfans-experiences-big-beyonce-bump-following-savage-remix} OnlyFans creators traverse the intersection of professional content creation---a growing form of gig work---and sex work. As a result, especially for those who were not previously working in the sex industry, joining OnlyFans to create sexual content can be uniquely stigmatizing. 
Our work aims to understand how those new to the commercial sex industry navigate content creation on OnlyFans and are motivated to overcome potential barriers to chose OnlyFans over other labor options. 
%

We find that celebrity hype around OnlyFans created by popular culture news such as Beyoncé mentioning the platform in her 2020 ``Savage'' Remix and creator Bella Thorne making $\$1$ million in her first day on the platform~\cite{sanchez2022world}, led the ``cultural assimilation'' of the platform, as one participant described it. This assimilation reduced the stigma of joining a platform associated with adult content and created significant interest among potential creators about how much money could be made. Further, the very design of the platform itself---which requires creators to advertise on mainstream social media platforms to drive traffic because there is no search or opportunity for creator discovery directly through OnlyFans---drastically raised the visibility of the platform, drawing in new creators. 

Beyond the hype and reduced stigma created by celebrity notoriety and the carefully curated visibility of the platform, creators were also motivated to join the platform by typical gig work motivations (e.g., money, flexibility, accessibility) and, similar to other fully-digital gig work, safety during the COVID-19 pandemic. However in many cases our participants considered creating on OnlyFans preferable to other gig work 
because of the platform's affordances: a lack of rating of creators, ease of blocking harassing subscribers, and autonomy over what content to create and how to construct their business model. Other factors that motivated creators were a desire to engage in sexual expression, the utility of the platform for curating their existing (non-commercial) sexual content, and the opportunity to leverage existing digital audiences or skills from other recreation or labor (e.g., cosplay, waitressing) in which they were already engaged. 

Finally, the pandemic resulted in both increased subscriber-side demand for digital entertainment and increased creator-side need for paid work due to job loss. We highlight how our participants intersect with and diverge from other workers in the spaces of professional content creation, the gig economy, and sex work. 
%

\section{Related Work}
\label{sec:related}
Here, we review prior literature on opportunities, risks, and challenges in digital content creation and gig work, as well as OnlyFans in particular.
%
%


\subsection{Content Creators}
Prior work has studied and investigated people's motivations to recreationally and professionally create digital content. Prior work finds that recreational creators have a variety of motivations, including a desire for self-expression, recognition, to promote ideas, to help or inspire others and/or to build community~\cite{brake2014we, craig201927,omar2020watch,markman2012doing,nardi2004we, leung2009user,kopf2020rewarding,torhonen2019fame,noonan2018social,weber2021s}. Creators may become professional creators -- that is, monetize their creation, by happenstance or more intentionally, driven by entrepreneurial spirit, a desire to generate their own ``media brands'', to influence others, or to become famous~\cite{giardino2021social,matikainen2015motivations,freberg2011social,de2020unravelling,noonan2018social}. 
Prior work on motivations behind online video content creation on services such as YouTube and Twitch suggests that enjoyment and socialisation are more significant drivers for content creation than income and reputation~\cite{torhonen2019fame}.
In contrast to platforms such as YouTube, creators on OnlyFans do not have the affordances to attract fans from the platform itself~\cite{van2021interdependent, safaee2021sex}.
In order to succeed, creators must promote themselves and their content on other social media platforms~\cite{bonifacio2021beyond}.
Hence, OnlyFans creators fall in the category of influencers (internet micro-celebrities) as defined by prior work since they \textit{manage and engage} with relatively large followings on social media platforms for \textit{self-branding and recognition}~\cite{khamis2017self, van2021interdependent}. 
In this work, we explore how the motivations of OnlyFans creators intersect with and diverge from those of other content creators.

\subsection{Gig Workers}
%
The online ``gig economy'' is defined as a labor market where independent contracting happens through, via, and on digital platforms~\cite{woodcock2019gig}.
A digital gig economy platform is one that offers tools to bring together the supply and demand for labor~\cite{graham2018towards}.
The work performed on gig economy platforms is non-permanent, and lacks job security~\cite{woodcock2019gig}. While gig work is often associated in the digital context with ridesharing and crowdwork, a far broader collection of workers engage in gig work. For example, scholars have shown that those working in the porn industry~\cite{safaee2021sex, van2021competing, easterbrook2022onlyfans, jones2020camming, butler2020aligned}, professional content creators more generally~\cite{vallas2020platforms} and care workers who find employment via online marketplaces~\cite{hunt2019women, kasliwal2020gender} are all gig workers.

According to the $2021$ Pew Research Center survey, $16\%$ of Americans have engaged in and earned money through online gig work
~\cite{
pewresearchsurvey2021}.
Several studies have analyzed the factors that influence the growing adoption of gig economy platforms, finding that this work offers greater autonomy and flexibility to clients and workers compared to traditional formal labour~\cite{woodcock2019gig, wood2019good, graham2017risks, lehdonvirta2018flexibility}. 
Further, gig workers may be able to access more clients and thus greater income through platforms that connect them with clients from diverse industries and in some cases allow them to take on increasingly complex tasks as they gain more experience~\cite{wood2019good, graham2017risks}. Prior work finds differences in the motivations of part- vs. full-time workers, such as ride-hailing drivers: Rosenblat
found that part-time ridehail drivers are mostly motivated by the flexibility of work offered by the platforms, whereas full-time drivers were motivated by having similar previous experience or lack of other job opportunities~\cite{rosenblat2016motivates}. 

Gig work platforms have low entry barriers along with flexible work hours and locations~\cite{rosenblat2016algorithmic}. As a result they may be more accessible to marginalized communities. Prior work has shown that gig work may offer greater ease of entry for disabled people and may provide necessary benefits such as  greater control on when and how tasks are performed compared to traditional employment~\cite{harpur2020gig, sannon2021workers, ali2011types}. Gig work may reduce/avoid the need to disclose non-task-relevant disabilities, thereby reducing potential stigma and bias~\cite{harpur2020gig}. Across American gig work platforms, women gig workers make up approximately half of the workforce~\cite{gigeconomystats}.
Prior work has focused on understanding platform engagement of women in the gig economy and exploring their challenges to economic opportunities and work-life balance~\cite{hunt2019women, raval2019making, chaudhary2020india}.
Through surveys and focus group discussions, Chaudhary finds that women are attracted to gig work because of flexible hours and easy management of other commitments, more income-generating potential and the opportunity to become potential breadwinners in the family~\cite{chaudhary2020india}.
Although flexibility is the key motivator among women, they had to make difficult trade-offs between their time-use, income generation and caring roles~\cite{hunt2019women}.
In a recent study~\cite{ma2022brush}, it was found that gig work platforms are gender-agnostic towards women's experiences and values and leave women vulnerable to bias and harassment by not enforcing anti-harassment policies in their design. This results in marginalization of workers as platforms discard the various social contexts that these workers are coming from~\cite{gray2019ghost}.

While gig work platforms offer some preferential working conditions over fixed labour jobs, prior work also finds that gig workers suffer from low pay, crowded marketplaces, discrimination, employment insecurity, social isolation, overwork, financial precarity, and stressful and dangerous working conditions~\cite{graham2018towards, graham2017risks, rosenblat2016algorithmic, donovan2016does, seetharaman2021delivery, stewart2017regulating}.
Gig workers do not enjoy the benefits and protections provided by labor and employment laws as well as other benefits such as paid sick leave, health insurance and pensions~\cite{donovan2016does, doucette2019dual} and are forced to take on additional costs such as damage to tools, injury costs, and unpaid gaps between between paid gigs~\cite{zwick2018welcome, vallas2020platforms}.
While gig workers experience increased flexibility by using gig work platforms, workers are dependent on those platforms for payments and jobs. This creates significant vulnerability as the risk of losing platform access (i.e., being deplatformed) is ever present~\cite{vallas2020platforms, pangrazio2021old}. Further, prior work~\cite{rivera2021want} suggests through qualitative surveys and semi-structured interviews that gig economy platforms lack potential for career development. As a result, gig workers must transition out of those platforms to achieve their long-term career goals. 

Finally, recent works have also studied the impact of COVID-19 on the gig economy, finding increases in opportunities for fully-digital gig work, decreases in the number of in-person gig workers, and changes on the part of in-person gig work platforms, including both positive (employment benefits, including limited paid sick leave) and negative changes (restricting the flexibility and freedom of workers)~\cite{umar2021impact, cano2021flexibility, katta2020dis,apouey2020gig,dunn2020motivation}.
%

\subsection{OnlyFans}
\label{sec:related:of}
OnlyFans creators are independent contractors and earn their money through the platform's business model of subscription and tips~\cite{safaee2021sex}, much like other gig economy workers.
In this work, we highlight similarities and differences between the motivations of those joining more typically studied gig work platforms such as those aforementioned and OnlyFans creators. Further, many of our participants had previously participated in other forms of gig work prior to joining OnlyFans and we report on their perspectives on the differences and similarities in their labor experiences.

Between the stigma of studying sex work and OnlyFans being a relatively new phenomenon, the academic research on this platform is still in its infancy, despite an enormous body of media coverage regarding creators, fans, changes in OnlyFans terms and conditions, and celebrities on the platform ~\cite{nytimes2019onlyfans, nytimes2021onlyfans}.\footnote{We have collected numerous additional media articles about OnlyFans here: \url{https://osf.io/ws54t/?view_only=3f553c40cfd940728f8e6329cd9a2795}.}
As a result, much  existing academic work on OnlyFans is presented in student capstone projects and theses rather than peer-reviewed papers. 

Offering empirical work on OnlyFans users broadly---including both creators and fans---Arañez Litam et al.\ recruited MTurk workers as well as university students who are OnlyFans users and compared their attitudes toward sex with those who were not on the platform, finding that they were similar in terms of attitudes towards sex~\cite{litam2022sexual}. They additionally studied the demographics of OnlyFans users in their sample, but do not distinguish between fans and creators; they report that most of their participants were heterosexual, white, married men, consistent with prior work on the consumers of digital sexual content~\cite{rissel2017profile, regnerus2016documenting, attwood2005people}. 
We answer their call to conduct qualitative research with Onlyfans users, concentrating specifically on creators new to content creation. As they propose, we directly examine the experiences of creating on the platform. 

Additional empirical prior work focuses on particular sub-populations of OnlyFans creators. Cardoso and Scarcelli carried out a qualitative research study with 20 creators, all young, white, Italian, cis women~\cite{cardoso2021bodies}. They contextualise OnlyFans by explaining the rise in popularity of amateur and gonzo\footnote{Gonzo is a stylistic and aesthetic genre of pornography characterised by its anti-thesis to prior big-budget studio porn, with POV (point of view) camera work and pseudo-documentary style narrative. See e.g. Porn Studies Issue 3:4, introduced by Biasin and Zecca~\cite{biasin2016introduction}.}
 porn genres. The study looked at the somatic experience of content creation and analysed the results by examining the kinds of corporeal production enacted by creators. Cardoso and Scarcelli begin to identify different categories of creators, distinguishing between creators who consider themselves full-time professionals vs. those who view their work more casually. Ryan studies male sex workers, finding that they use OnlyFans as an extension of their work as micro-celebrities on Instagram in their gay communities~\cite{ryan2019netporn}. Hamilton et al. investigated sex workers who pivoted from in-person to online work during the pandemic~\cite{hamilton2022risk}, finding both new benefits and harms from this new modality. Our participants therefore form an additional hitherto unstudied population. 

Additionally, three smaller scale projects published as posters or theses empirically study OnlyFans creators. Ebersole interviewed eight undergraduate women who are OnlyFans creators, finding that they were working on OnlyFans for additional income; that some workers used the platform to express their sexuality; and that significant labour was also carried out off-platform and that creators connected in order to support their work, themselves and each other~\cite{ebersole2022online}. Uttarapong et al. also focused on this community aspect, interviewing 15 creators to find out how they used their personal networks for business and also for support~\cite{uttarapong2022social}. Our work includes creators with vastly bigger audiences than their largest, who although not being celebrities, still utilized the intra-community practices described in their paper (among others) to great effect. Finally, Dominguez’ thesis, a qualitative study (n=6) also studied creators' support networks~\cite{barroso2022redes}. 
%
In theoretical analyses, multiple scholars note how neoliberal discourses\footnote{Neoliberalism is the political and structural phenomena by which socialist policies and governance are systematically dismantled in favour of free-market individualism~\cite{bockman2013neoliberalism}.} set the scene for the rise of the self-employed porn performers on digital platforms 
and factor into the popularity of OnlyFans as a gig-work platform~\cite{easterbrook2022onlyfans, ryan2019netporn}. 
They conclude that digital sex workers are more precarious than other gig workers because of the additional sex work stigma, related regulations (e.g., FOSTA-SESTA), and resulting structural implications (e.g., restricted banking access). 
Lastly, prior work suggests that other sex workers build communities online for activism, networking, peer-support, and sharing health and safety information~\cite{sanders2018internet, barwulor2021disadvantaged, bernier2021use, mcdonald2021s, barakat2022community, strohmayer2019technologies}. 

Outside of the specific study of OnlyFans, there exists a large scholarship focused on many kinds of sex work including pornography and sex work that is digitally mediated. Whilst a full examination of all of these areas would be impossible within the scope of this paper, we highlight the findings of a subset of these works that are particularly relevant to our context. 

Much of the scholarship regarding online sex work involves the study of web-camming, which takes the form of one-to-one or one-to-many live-streaming shows that may involve stripping or sexual activities. Jones undertook a large-scale study of webcam workers~\cite{jones2020camming}, including particularly investigating racism~\cite{jones2015black} in the industry and the experiences of transmasculine~\cite{jones2021cumming} and fat~\cite{jones2019pleasures} cam performers. Jones' work clearly shows that intersectionality affects both experience and earnings in digital sex work. Examining in-person work, Berg interviews more than 80 porn workers and, in analyzing these interviews through 
a Marxist-feminist labor-studies lens, finds her participants are undertaking ``pleasurable resistance'' to other kinds of labor~\cite{berg2021porn}. Perhaps in an effort to capitalize on this sentiment, in analyzing the terms and conditions of webcam sites, Stegeman concluded that webcamming sites attempt to undermine the labor rights of performers by framing camming as ``not-work''~\cite{stegeman2021regulating}. 


This dependence on both sex work (e.g., webcamming) platforms and on other platforms to engage in labor makes deplatforming (having an account banned or removed) a particularly onerous risk for digitally-mediated sex work. Prior work on digital marginalization of sex workers shows that platforms of all kinds regularly deplatform sex workers, taking away opportunities not only for business but for social participation and peer-support by censoring conversations around sex work and even sex education~\cite{barakat2022community,barwulor2021disadvantaged, blunt2020erased}.
Blunt et al.\ discuss how online sex-work platforms constantly update their algorithms and terms of services to sanitize online space, impacting sexual expression while simultaneously monetizing the data and content of online sex workers~\cite{blunt2021automating}.

Finally, more broadly examining ``sexual entrepreneurship,'' Rand describes digital sex work as absent from literature on digital labor~\cite{rand2019challenging} and argues for the inclusion of sex workers in academic and other conversations on gig work and digital labor. 
%
We answer this call and build on the above-mentioned prior work---focused on provision, benefits, and challenges of digitally-mediated sex work---by exploring the motivations of adults who do not have prior sex-work experience to create and monetize adult content as a form of gig work on a relatively mainstream subscription-based social media platform: OnlyFans.

\section{Methodology}
To understand the experiences and motivations of those new to professional sexual content creation to start a subscription service on OnlyFans, we conducted $22$ semi-structured interviews with current and past OnlyFans creators in the United States during September and October $2021$ during the global COVID-19 pandemic.

\subsection{Participant Recruitment}
We sought to recruit a broad range of U.S.-based content creators on OnlyFans. We intentionally avoided references to sexual content in our recruitment materials (see \url{https://osf.io/ws54t/?view_only=3f553c40cfd940728f8e6329cd9a2795} for anonymized recruitment graphic) and interview protocols, in order to be inclusive of participants making any type of content. Despite this, all of our $22$ participants did create sexual content on OnlyFans. We recruited our participants online by posting on social media and sending out recruitment flyers to the researchers' contacts to share in their networks~\cite{biernacki1981snowball}, and in-person by posting flyers around several university campuses and cities (at coffee shops and grocery stores) in different parts of the U.S. 

The recruitment materials linked to a Qualtrics screening survey. OnlyFans does not publish demographic data about their creators~\cite{litam2022sexual}, and our goal in sampling was not to interview a representative sample of creators; rather we sought to hear from a diverse set of creators, including and especially from those with identities that might impact their experience of the platform. To do this, we intentionally set quotas for combinations of race, gender and age to the sign-up process in Qualtrics in an attempt to ensure that we recruited a diverse range of participants. We also used the survey to screen out those who had previously created content on OnlyFans but had stopped creating such content more than 18 months ago (as these creators would be less familiar with the current platform affordances and updates), and those who had previously engaged in sex work, as the focus of our work was to understand the motivations and experiences of those creators who were new to sex industry work (if that was the type of content they were creating on OnlyFans) as they may face unique barriers and challenges in their labor (see Section~\ref{sec:related:of} for a summary of prior work on the experiences of those who have prior sex work experiences on OnlyFans). 
Qualified participants were redirected from the screening survey to anonymously self-schedule for an interview using a Calendly link.

\medskip
\noindent\textbf{Participant Demographics.}
All of our participants were $18$ years old or older, with median age of $29.5$ years ($\sigma=7.01$).
Participants were invited to self-describe their gender and race/ethnicity in our screening survey. As such, our participants self reported their genders as: woman $(12)$, man $(3)$, non-binary $(8)$, and three participants self-described as trans.\footnote{We report gender in line with guidance from \url{https://www.morgan-klaus.com/gender-guidelines.html}.}
Participants who reported their race/ethnicity reported as: Black $(3)$, Asian $(2)$, Asian and white $(1)$, Hispanic and white $(2)$, and white alone $(11)$. 
$9$ out of $22$ of our participants reported that they had a disability. 
Our participants were also asked their education status in the screening survey, reporting their highest level of education as a(n): Advanced degree ($2$), Bachelor's degree ($4$), Associate's degree ($3$), some college coursework but no degree ($8$), or a High School diploma ($2$).

$10$ of our participants reported in the survey that they had done gig work previously, on food delivery platforms (UberEats, PostMates, GrubHub, DoorDash), service provision websites (NextDoor, Craigslist, Facebook Marketplace), or ride-hailing platforms (Uber and Lyft). $6$ participants had done other paid digital work such as tutoring, marketing, selling art, or writing reviews. Lastly, three of our participants described \emph{during} the interview that they had participated in some kind of sex work prior: we had attempted to only interview those without such experience but can account for this by understanding that these participants could have either mis-read the screening survey or not perceived their work (stripping, phone sex, professional BDSM work respectively) as sex work. 

Our participants had spent a median of $15.5$ months using OnlyFans; three of our participants had stopped using OnlyFans less than $6$ months prior to the interview.
Our participants reflect a wide range of income and subscriber numbers: small creators and creators with enormous reach. We asked the total income earned to date and found (in U.S.\ dollars) the minimum was $\$135$ and the maximum $\$332,000$ with a median of $\$1,237.84$ ($\sigma=\$83,974.24$). The minimum number of subscribers reported (at the time of the survey) was $10$ and the maximum $13,000$ with a median of $60$ ($\sigma=4,143.13$). For seven of our participants, OnlyFans was their primary source of income. Lastly, our participants also reported their popularity ranking on the OnlyFans platform (a statistic provided by the platform); they provided the rank of their highest page if they had multiple pages. The most popular among our participants was in the top $0.03\%$ of creators and least popular was in top $76\%$ with a median popularity of $8.7\%$ ($\sigma=24.82\%$). 

\subsection{Interview Data Collection}
We used semi-structured interviewing methodology~\cite{bernard2013social}. 
%
%
Interviewers first explained the goals of the study and answered any questions that the participants had before requesting to record the interview.
At the start of the interview, participants verbally gave their consent to be interviewed and recorded and interviewers re-confirmed with each interviewee the screening survey information that was filled out by the participant. 
Participants were then asked background questions to understand the kind of work they did before joining OnlyFans. We then asked questions related to their experiences in those prior jobs and asked our participants to make comparisons with their experiences on OnlyFans (e.g., how do you find OnlyFans different or similar to these prior/current experiences?).
We asked our participants questions about their initial exposure to OnlyFans and their motivations to start creating, publishing, and promoting content.
The interview then queried what kinds of content the creators were making on OnlyFans, and the kinds of skills or interests they had to do this work.
Finally, we asked our participants' platform-related questions, which are out of scope for this paper. 
Participants were also invited to not answer any question that they did not want to answer.
The relevant interview questions can be found at the following anonymous repository: \url{https://osf.io/ws54t/?view_only=3f553c40cfd940728f8e6329cd9a2795}.

All the interviews were conducted in English via chat, voice, or video.
Each interview lasted between $37$ and $96$ minutes, with an average interview length of $60.5$ minutes. Participants were compensated with \$50 via Amazon gift card or PayPal.
All the interviews were then transcribed and any incidental identifying information was removed.

\subsection{Data Analysis}
We analyzed the interview data using an iterative open-coding process~\cite{strauss1998basics}. One author randomly selected five interview transcripts to identify common themes and create a thematic framework for the interview data. After creating the codebook, two of the authors coded another five transcripts. 
The coders achieved a Cohen Kappa of $0.75$, which is considered a substantial inter-coder reliability score~\cite{landis1977measurement}. Subsequently, both the coders resolved any minor disagreements and refined the codebook, which was then used by one of the authors to code all the remaining transcripts.
Given the qualitative nature of our results we report our findings primarily qualitatively, sparingly providing counts of overall themes (i.e., how many participants reported each motivation) only to highlight the prevalence of patterns in the interview data and provide transparency into data analysis and theme emergence. 
Moreover, we want to highlight that no inferences should be made about the prevalence of themes observed beyond the sample given the qualitative nature of our sample~\cite{mcdonald2019reliability}.

\subsection{Ethics, Impacts \& Research Justice}
Our work was approved by our institution's ethics review board.
Since our participants are 
undertaking stigmatized work, we took extensive measures to protect participant data, anonymity, and privacy at every stage.
Some examples include using platforms that are end-to-end encrypted for conducting virtual interviews (e.g., paid Webex), offering instructions on how to create an encrypted (ProtonMail) email address to organize an interview and receive reminders, and using a scheduling system (Calendly) that did not require personal information such as name from participants for scheduling. We offered participants multiple options for compensation: anonymous Amazon gift cards or PayPal payments (which could be sent to an encrypted email address they had created using our instructions). We took great care to protect the identities of participants and any unintended identifying information has been removed. 

In alignment with research justice principles, we employed a sex worker to transcribe the interviews and will send a copy of published research using this data to those participants who requested it~\cite{macneil2006informing,bhalerao2022ethical}. Lastly, as detailed further below in Section~\ref{sec:meth:positionality}, we consulted OnlyFans creators on our research design and interview protocol and they voluntarily circulated the study recruitment materials in the networks of creators that they were members of; these creators were compensated for this labor.

\medskip
\noindent\textbf{Broader Perspectives.}
Our hope for the positive impacts of this work is that adult content creators are taken seriously and considered part of the gig economy workforce. 
There have already been efforts to use research on gig workers in attempts to improve their working conditions and regulate employers and platforms in this area to their advantage (see the many papers in Section~\ref{sec:related} for research on gig workers). However, the stigma of sex industry work often prevents adult content creators from being included in these efforts, which leaves them behind in terms of workers rights~\cite{butler2020aligned}. Sex worker and gig worker rights groups can use our research in conversation with policy makers to improve the working conditions for all content creators. 
We will provide plain-English, 1-page summaries of the results of this research project to be disseminated widely among the population studied. OnlyFans creators are already in community and overlap with sex worker communities, with whom we already have established relationships. 
Potential harms from the publication of this work include use of any negative experiences reported in our paper by anti-porn and anti-sex-work lobbying groups to advocate for OnlyFans banning adult content, as there has already been pressure to do~\cite{van2021competing}. 
 
\subsection{Positionality}
\label{sec:meth:positionality}
The researchers are scholars of technology and sex work. We consulted with OnlyFans creators on the research design, interview protocol, and recruitment. However, those creators were not directly involved with the data analysis or writing of this paper. Additionally, there were participants with genders and ethnicities who are not reflected in the research or interview team, notably male and Black creators, since we are white/brown and American/british\footnote{Authors prefer not to capitalize british as a decolonisation practice~\cite{website:mtroyal}.}/Indian, which may limit our capacity to fully understand and analyze the nuance of that data.

\subsection{Limitations}
%
Our study has several limitations. First, we conducted interviews only in English and only of U.S.-based creators, which limits participation to U.S.\ English-speakers. Future work should explore the motivations and experiences of OnlyFans creators in other geographies and contexts. Second, while we aimed to recruit only OnlyFans creators who had not done other forms of sex work previously, and used our screening questionnaire to screen for these participants, three of our participants had previously done sex work (see Participant Demographics above). Third, our interviews ask participants about their experiences conducting business in a competitive market. Thus, it is possible participants omitted relevant information about their business strategies.

\section{Results}
Overall, we find that our participants were motivated to join OnlyFans due to: 1) the visibility and acceptance of OnlyFans in society, which was created through a combination of celebrity hype, platform design, and peer conversation; 2) the potential earnings on OnlyFans, which they perceived as a (better) alternative to other forms of gig and service work; 3) a desire to engage in digital sexual expression; 4) already having existing content, audiences or skills that allowed them to quickly build a profitable OnlyFans page; 
and 5) pandemic factors such as increased flexible time, increased safety concerns about other forms of gig work, and loss of other forms of income due to economic impacts.

\subsection{Visible \& Accepted: OnlyFans in Society}
\label{sec:visibility}
OnlyFans gained significant visibility and mainstream acceptance, particularly starting in 2020. This came as a result of: 1) celebrity participation and conversation about OnlyFans, 2) peer conversation about the platform as a result of reduced stigma from (1), and 3) high visibility of the platform from both (1), (2) and OnlyFans' distinct design: it offers very limited creator discovery on the platform, pushing creators to promote their content and the platform itself across mainstream social media platforms.

All 22 participants discussed this visibility or acceptance in some way. Our participants felt comfortable becoming OnlyFans content creators as a result of the societal pervasiveness of OnlyFans combined with broad participation from celebrities and peers, which decreased the stigma of creating on OnlyFans. For example, P11 explained that COVID-19 and the positioning of OnlyFans as ``part of the gig economy'' made people more comfortable with consuming and creating adult content through the platform. Thus, ``every day regular people are associated with'' OnlyFans, which made it feel like the platform had a ``personalized feeling to it of like this isn't for porn, this is [a platform] for someone I just happen to know [to create adult content].'' As a result, P11 explained, there is ``less of a whorephobic\footnote{https://www.theguardian.com/commentisfree/2010/jun/23/sex-workers-whorephobia} stigma'' around creating content on OnlyFans. Because of the growth of OnlyFans and lessened stigma around the platform, P5 remarked that ``in the future I think that there will be a lot more acceptance of sex work'' in general, not just as part of OnlyFans. 

\medskip
\noindent\textbf{Celebrity Hype.} Celebrities announcing on their social media and in the news that they were starting OnlyFans accounts also helped bring attention to the platform. Cardi B joined in August 2020\footnote{https://www.rollingstone.com/music/music-news/cardi-b-onlyfans-1043254/} after, as aforementioned, Beyoncé mentioned the site in a remix in April causing, according to a spokesperson for OnlyFans, ``a $15$ percent spike in traffic'' in the subsequent $24$ hours.\footnote{https://www.thedailybeast.com/adult-site-onlyfans-experiences-big-beyonce-bump-following-savage-remix} It is important to note that clip sites (websites where adult performers can upload video content for clients to purchase) have existed for a long time, as have sites with other features that OnlyFans has, such as live streaming.  
However, OnlyFans' specific brand notoriety was important for our participants; $12$ of our participants described OnlyFans as ``compelling enough'' to try out because of the promising, lucrative self-employment enabled by the platform. For example, P2 described OnlyFans as ``becoming a mainstream term and semi-accepted (more so than other similar sites)'' and P8 told us that the mention of OnlyFans by ``internet celebrities...was definitely how it caught my attention first.'' However, it was not simply finding out about OnlyFans through celebrity mentions, but also imagining why celebrities joined the platform that inspired our participants: P8 continued ``I thought it was cool, I liked the idea of these mostly women... reforming sexuality in their own way and being able to claim, reclaim something.''


Interestingly our participants P5, P10, and P21, were also motivated 
by the monetary hype created by celebrity earnings on the platform. The high earnings of several specific celebrities were referenced by our participants as a motivating factor to join.
P5 discussed Bella Thorne joining OnlyFans, remarking ``she's already fairly famous, but then she did one day of OnlyFans and she made one million dollars...that was pretty persuasive.'' Similarly, P10 ``was swayed to do it [OnlyFans] after the Bella Thorne incident.''

Moreover, four of our participants (P1, P2, P10, and P15) read articles about how non-celebrities had increased their income by changing from regular jobs to OnlyFans work: ``there's a nurse that started OnlyFans after the pandemic started and started making more money than being a nurse.'' (P1). P2 had ``seen a couple posts about people being able to buy a house from OF earnings.'' 

Overall the celebrity allure, with the promise of money, was a common theme. P17 explains how these threads combined: ``at first I was hearing that it was lucrative...[and then] I kept hearing about [OnlyFans everywhere]...[I] wanted to see what the hype is about.''

\medskip
\noindent\textbf{Peer Suggestion.} Perhaps as a result of the reduced stigma created by celebrity discussion of and participation on OnlyFans, we observed that a significant number ($12$) of our participants' friends and family also directly discussed OnlyFans with them either as a conversational topic, or because they were creators themselves. P20 explained that they were motivated to join OnlyFans because a friend told them about it, saying ``oh you should probably just start OnlyFans and [they] started sending me [clients].'' Additionally, these current creators could serve as information bridges, helping new creators onboard. For example, P14 described how her sister encouraged her to join and collaborated with her: ``Yeah my sister, she's like you have a great body why don't you just try OnlyFans? I was only going to do it for maybe just a couple of weeks just to help take care of things financially, but it lasted longer than that.'' 
P13 was similarly encouraged by the experiences of others: ``I also had a friend who started doing it...so I started talking to her about it a little more and asked her how is it and she had nothing but good things to say so I was like okay I'll try it out.''

\medskip
\noindent\textbf{Platform Design.}
Finally, $16$ of our participants described finding out about OnlyFans through other social media sites, where they saw friends or influencers posting about it. P8 describes how ubiquitous mentions of OnlyFans became mainstream on non-sexual content platforms: they ``found out about [OnlyFans] through Twitter'' and ``had seen a few comments on it and it was trending at one point.''

Since OnlyFans does not have internal search functionality nor robust content discovery, creators have no choice but to use external advertising to drive clients to their accounts. In turn, this also amplifies the popularity of OnlyFans to other potential creators. P1 told us that ``on every big tweet that blows up, there's usually someone's OnlyFans invite.'' Out of the $16$ participants that described learning about OnlyFans through social media sites, more than half ($9$) mentioned Twitter. P12, who called the phenomenon of becoming aware of OnlyFans ``cultural assimilation,'' added that ``many people who I know and followed on Twitter were talking about and in some cases had OnlyFans accounts.'' One of the major reasons our participants mentioned Twitter more than any other mainstream social media sites is because Twitter allows consensually produced adult content provided the media within the Tweet is marked as sensitive.\footnote{https://help.twitter.com/en/rules-and-policies/media-policy}

\medskip
\noindent\subsection{OnlyFans as Work}
\label{sec:results:ofwork}
$18$ of our participants mentioned money as a motivator to create on OnlyFans. While for three (P1, P6, P13) of our participants this was the only reason they created an OnlyFans---as P1 put it ``money! Honestly, that's about it''---for the rest earning money was one of multiple reasons for joining the platform as a creator.

Along with the known motivations---flexibility, autonomy, control, income---of gig workers to join gig economy platforms~\cite{rosenblat2016motivates,chaudhary2020india,wood2019good}, our participants perceived OnlyFans as better than other forms of gig work either due to 
the accessibility of the platform, more control, more income or replacement of lost income, enjoyment, ease of boundary setting, and physical and pandemic safety. It also allowed them to attend to increased responsibilities as carers or parents. 
%
%

\medskip
\noindent\textbf{Similarities to Other Work.} 
Similar to other gig workers, $15$ of our participants sought the autonomy to direct their own working hours on OnlyFans. P12, comparing it to other jobs, described it as ``one of the better ones that I've had, just in terms of affording me flexibility.'' P14 summarized it thus: ``I can sign in on my own times, my own hours... the fact that I could schedule my own time, I don't have a set time... with OnlyFans I can just show up.'' This autonomy was even described as pleasurable for some, with P10 explaining: ``OnlyFans allows me to keep a schedule that I enjoy.''

Beyond the flexibility of being able to choose their own hours, there were other kinds of accessibility advantages that participants described. P4, who had multiple health diagnoses, had delivered food prior to working on OnlyFans; however: ``I can still do OnlyFans even if my legs aren't working...because I can just kind of sit.'' Neurodivergence was also mentioned by two of our participants as a reason why OnlyFans was a more accessible workplace than regular jobs. For example, P21 explained that ``due to the nature of my ADHD, my neurochemistry makes it very hard to maintain schedules, routines, consistency.'' P16 added ``I have a lot of trouble working like regular jobs,'' describing their neurodivergence as ``a bit of a barrier'' to other kinds of work. Johnson has documented similar accessibility in digital patronage work for disabled creators on Twitch~\cite{johnson2019inclusion}.

The flexibility to attend to caring responsibilities also made OnlyFans a viable workplace for two (P4, P20) of our participants. Even though P20 had already been working from home, when their mother became ill, they lost that job due to workplace surveillance: ``they're watching you, if you leave the computer they know that you left the computer so it was one of those things where I couldn't work and take care of her.'' 

\medskip
\noindent\textbf{Better than Other Gig Work.} 
Whilst most of our participants did not make the kind of money they described hearing about in the discourse that motivated them to join (see Section~\ref{sec:visibility}), six of our participants (P2, P5, P7, P14, P19, P20) did experience increased earnings compared to other informal or gig work. For example, P2 described ``making like 25 dollars for a night of pet-sitting where I had to stay at someone's house vs I can make that amount with 1-2 clip sales.'' Similarly, P5 felt that work on OnlyFans is ``pretty easy... I do it maybe 10 minutes a day, it doesn't feel like a job.'' Further, OnlyFans creators control the price of the content and services they offer, not the platform. When asked to compare OnlyFans with other gig work, P2 went on to say: ``I make more money on OF...And I can control the price more.'' 
%

Three participants also described OnlyFans as being better than other gig work not only in terms of earnings, but in terms of their power in the labor relationship. Prior work~\cite{kim2018impacts, raval2016standing} describes the additional emotional labor gig workers must expend in order to receive good reviews, which helps their algorithmic performance. However, on OnlyFans, creators are not reviewed. P2 described that because ``there isn't a rating system like in so many app based gigs,'' on OnlyFans it was ``easier to draw boundaries''. Even in comparison to face-to-face work, P21 explained that OnlyFans offers more autonomy and thus: ``allowed for me to establish my boundaries better, unlike the customer service jobs I've worked before where I can't talk back, I can't cut out a customer because it's up to the manager, whereas [on OnlyFans] I have the control to just block someone.'' P13 agreed, describing that in food service jobs they had been ``hit on relentlessly,'' and that ``the biggest difference'' with doing OnlyFans compared to other work ``has been respect.'' 

Finally, three (P5, P6, P7) of our participants preferred OnlyFans to other forms of gig work or service work because of the literal physical distance it offers from clients. P7 explains: ``I definitely make more money on OnlyFans than I did with Uber and Lyft and I don't have to drive around all day and I feel safer because I'm not actually meeting with people. Before, strangers were getting into my car. It was always fine, but now there are no strangers that I'm seeing.'' Much like those who had done sex work prior to the pandemic~\cite{hamilton2022risk}, our participants used OnlyFans to avoid face-to-face contact and the risks associated with meeting strangers in person during this time. 

\medskip
\noindent\textbf{Unique Challenges of OnlyFans.} 
Although joining OnlyFans was lucrative and promising in the success stories our participants heard, many shared that overall OnlyFans earned them less money ($6$) and required more effort ($17$) than they anticipated. Moreover, while $13$ of our participants reported exclusively positive experiences with the platform and found the work to be better than other types of gig work, the others shared both positive experiences and challenges. 
OnlyFans requires significant marketing and promotion (P1, P9, P20), offers no protection from chargebacks (P19), and is less socially acceptable than other gig work (P9, P21). Most notably, because OnlyFans does not match creators to clients like many other gig work platforms or offer internal creator listing or promotion features, creators must work to draw fans from other platforms. This contributes to the mainstreaming of OnlyFans but can also be one of the more challenging parts of being a creator. For P9, ``promoting is definitely the most time consuming of all of it... people have to know it's out there in order to buy it.'' P20 said they find themselves doing OnlyFans work ``all day every day... I spend hours hours hours, 17 hours on Telegram the past 24 hours, eight hours on Safari, most of that is OnlyFans, five hours on Twitter. Like it's just, it's obscene.
''

Furthermore, our participants described multiple factors that could impact their success on the platform, including the level of explicitness of their content and their gender/racial identity. Some of our participants expressed a perception that there is a hierarchy of OnlyFans creators based on: the type of content (explicit content creators were perceived to be more successful), the type of fan base (celebrities/famous porn performers perceived to be more successful), the effort expended promoting the content off-platform, and lastly the sociodemographics of the creator. 
Regarding the latter, our participants specifically suggested that white women or those with an existing audience tend to have an easier time on OnlyFans, whereas fat, disabled and racialized performers as well as straight men might have greater struggles building a following on OnlyFans and making ends meet. P20 accurately summarizes the aforementioned collective observation by mentioning that ``intersectionality is a bitch'' and ``race and body type has a lot to do with it (how much you earn on OnlyFans)''. They ended their response by adding that ``if you're disabled, if you're fat, if you're poor because being poor affects every stuff too, that affects your content production, that may even affect your ability to even engage your fans.''

Finally, the popularity of OnlyFans means that much of the sexual content market has moved to the platform. Even participants who want to use platforms other than OnlyFans felt they had no choice if they wanted to be successful. For example, P21 said, ``OnlyFans is so big that whether you like it or not, if you do online [sex] work you're gonna have to use that site to be able to live off of your income.'' P12 shared that they don't like OnlyFans as a company, but stay on the platform because ``it is where the audience is.''

\subsection{Sexual Expression and Community} Prior work suggests that more than $85\%$ of adults create and share sexual content as text messages, images, and videos~\cite{stasko2015sexting}. Participants in our study were no exception. Prior to joining OnlyFans they were already creating and sharing sexual content; $14$ of our participants described a desire to engage in sexual expression as a motivating factor for joining the platform. 

For these participants, OnlyFans afforded an opportunity to organize and monetize their content in a way that had not been available to them before. P3 described OnlyFans as a kind of sexual content archive, saying ``it's a good way to archive all of my stuff and have it in one place.'' P3 further explained that the ``general paywall'' on OnlyFans offered both privacy and an opportunity to financially benefit from content creation they already engaged in, saying, ``I've always been comfortable with showing myself off so [OnlyFans] was a good outlet for me to do that but also get a little benefit.'' P6 similarly described the situation of deciding to paywall their content: ``I was giving my nudes away for free and I was like why? Shouldn't I charge for these? So that's how I joined.'' Finally P2, who enjoyed creating pinup content said that OnlyFans offered ``a way to finally actually monetize it in a feasible way.''

OnlyFans also allowed participants to find a community of others who were open in their sexual expression. For example, P16 told us that OnlyFans ``was a way for me to explore my sexuality a bit and it seemed like a community that like might be positive for me.'' P22 went further, saying ``I envision a world where people are more comfortable with their bodies and with their own sexual expression, where the taboo of sex has been lifted... there's a part of me that wants to live those values, not just talk about them.'' Regarding colleagues on OnlyFans, P22 continued; ``being in community with other creators has been such a frankly unexpected joy.'' 

\subsection{Re-purposing Existing Content, Audiences \& Skills} $16$ of our participants were drawn to join OnlyFans because they had existing content, audiences or skills that they could leverage to quickly develop a profitable OnlyFans page.

As aforementioned, six of our participants were already creating sexual content recreationally. As a result, they had suitable content readily available to monetize and could make money instantly. P8 describes ``my first 70 posts that I made were previous and old photos that I had'' and that it was possible to ``make money off of something that I was already doing.'' 

Three of our cosplayer participants (those who dress up, typically as characters from works of fiction) decided to make their OnlyFans content cosplay themed in order to leverage their existing audience. 
For example, P2 was ``low key cosplay famous back in college'' and knew that they could exploit this ``(horny) fan base'' on OnlyFans. P13 had a similar motivation: ``I was already doing cosplay and I saw a lot of other cosplayers were having success with [OnlyFans] so I figured I would also try.'' Intersecting with a desire for sexual expression, P4 decided to use OnlyFans to showcase the ``different ideas or things that I had in mind but I couldn't post on other socials'' that don't allow ``18+ cosplay content.'' Much like P2, prior to starting their OnlyFans, P4 ``had people who showed interest in it [erotic cosplay content]'' and therefore felt there could be financial potential of creating this material to sell on OnlyFans. Rouse and Salter note that cosplay was already highly sexualized and that cosplayers use a variety of monetization avenues, including OnlyFans~\cite{rouse2021cosplay}. 

Another participant, P12, had already written a book about their sex life that had developed their ``reputation as someone who had a lot of sex'' (P12) and they opened an OnlyFans to give their followers ``a look into this thing [their sex life] that they were clearly already interested in.''

Beyond repurposing existing digital audiences, 
six of our participants described how their prior job experiences (tutoring-P1, webcamming-P5, burlesque-P7, waitressing-P9, modelling-P10, P18) offered transferable skills for OnlyFans creative work. For example P9, who had been a waitress, explained that ``waitressing is somewhat a show... you have to be very customer service friendly and you have to put on a little bit of an act to always be friendly because no one's always friendly 100 per cent of the time'' concluding that ``I guess working in the food industry and knowing how to please people is similar.'' P1, who was doing tutoring and OnlyFans at the same time explained: ``it's just all kind of tailoring how you present yourself to these people. You know I'm directly pulling tricks that I do tutoring over to OnlyFans... I'm kind of practicing the way I talk on OnlyFans and tutoring and vice versa.''

Finally, four (P12, P14, P16, P19) of our participants described an intrinsic sense that they would gain pleasure or be good at ``doing'' OnlyFans. For example, P14 said simply ``I knew that I would be good at it'' and P16 similarly explained that OnlyFans: ``just kind of seemed like something that I would enjoy doing.''

%

\subsection{Pandemic} $12$ participants described the COVID-19 pandemic as a motivator to joining OnlyFans. First, the pandemic decreased the appeal of in-person gig work (much like the decreased appeal of in-person sex work~\cite{hamilton2022risk}) due to the risk of catching COVID-19. For example, P7 explained that they had previously done ride-sharing work and had ``stopped because of the pandemic. So I was not going to have people in the car. Even now, I won't go back until we are way past this.'' Similarly P10 said ``I really needed work and I wanted to do something where I could work from home and not be exposed to COVID-19.''

Second, four (P5, P9, P10, P12) participants lost their jobs due to the pandemic and tried OnlyFans whilst waiting for unemployment benefits. P5 explained that ``when COVID-19 struck I had no income for a little while before I started getting unemployment.'' These factors (risk of catching COVID-19, loss of job, and lack of work available) combined powerfully for P9 who described: ``I was on unemployment after I lost my job for a little bit and then the benefits were coming to an end... I was a waitress before and I did DoorDash as well but I wasn't ready to go back to being a waitress... no one was going out to eat yet [because] it was still very not safe, so I need to make money somehow.''

Third, quarantines and lock-downs significantly changed how people spend their time. For example, as nearly all socialization and entertainment moved online, P22 described how ``there was a time around the start of the pandemic where especially a lot of gay men were creating private Twitter accounts to share nude photos. [I decided that] I should be making money for this because my body is valuable and if I can support myself by creating content that is really exciting.'' 

While some people---marginalized groups in particular---experienced an erosion of free time during the pandemic~\cite{kantamneni2020impact}, on average Americans reported more leisure time~\cite{bureau2020american}. Correspondingly, P17 described having ``nothing but free time'' and asked ``what's better than being at home in my room getting paid to be at home?'' On the consumer side, increased time at home, boredom, and loneliness during the pandemic spurred growth in OnlyFans and increased pornography use in general~\cite{zattoni2021impact, lau2021impact, grubbs2022porndemic}. P3 summarized this effect, saying, ``once the pandemic started and everyone was home [OnlyFans] definitely blew up.''
\section{Discussion}

OnlyFans provides many of the same opportunities for creators that other gig work platforms do. For example, flexibility allowed disabled participants and those caring for dependents to build schedules that suited them better than those imposed by regular work. Similar to other content creators, our participants also found OnlyFans to offer a creative outlet for (sexual) self expression. Beyond these factors, our participants were also attracted to join OnlyFans for a unique set of reasons specific to the platform, namely the brand notoriety and market share directly caused by two factors: 1) the celebrity hype and general public discourse surrounding the platform, and 2) the lack of internal discoverability, which forces creators to do external promotion. We posit that these factors also drove an increase in mainstream general acceptance of OnlyFans as a platform and reduced the stigma of participating in it as a creator relative to other forms of sex work such as webcamming and traditional pornography.

\subsection{OnlyFans as ``Playbour''}
As compared to other forms of gig work, the design of the OnlyFans platform has multiple factors that participants found preferable to other forms of gig work. First, whilst there is a black-box ranking system for creators to help convey how popular they are, OnlyFans has no direct rating system for creators. This is a distinction our participants found preferable to other kinds of gig work, which may subject them directly to client ratings and algorithmically dictate their ability to work based on those ratings~\cite{kim2018impacts, raval2016standing}. 

Second, OnlyFans allows for passive income opportunities. There is a hard limit to the number of jobs one may physically complete on Uber, Upwork, TaskRabbit and so on, which in turn limits one's possible income. On OnlyFans, the same piece of content can be sold multiple times, offering the potential
for greater financial return on the time/energy invested into it by the creator. 
The financial possibilities of OnlyFans over other types of gig work were a motivating factor for most of our participants despite the fact that many participants did not make the income they hoped. 
We note that there are several reasons why it is complex to study the actual per-hour return on work on OnlyFans: creators do many different kinds of work for their wage---including networking, promotion off-platform, and content creation itself---and different creators get different returns on their investments as a result of a variety of factors detailed in Section~\ref{sec:results:ofwork}. 
However, we interviewed more than one creator who had achieved enough steady content sales to generate substantial income although this had taken enormous amounts of persistent, creative, driven labor over many months. Further research should investigate the effort-to-income ratio---and the role of intersectional marginalization in this financial return---on OnlyFans in order to better support creators.
%

Third, the work itself could be enjoyable on OnlyFans. Whilst workers in the gig economy might find gig work \emph{preferable} to regular work, finding the work \emph{pleasurable} is relatively unusual. More than one of our participants described a sense of enjoyment in the work, in addition to those looking to fulfill their sexual expression needs by sharing erotic content. In \textit{Porn Work}, Heather Berg calls the participation in studio porn ``pleasurable resistance'' against the kinds of poor labor conditions found in regular jobs~\cite{berg2021porn}. Whilst some of our participants may not have joined OnlyFans if they had not needed the income, for many their content creation work was a pleasurable resistance against unemployment. Our participants are in many ways similar to other social media influencers~\cite{giardino2021social, matikainen2015motivations}, using sexual content creation as a kind of authentic self-expression and a vehicle for creativity. This is most similar to the concept of ``playbour''~\cite{torhonen2019fame}, where enjoyable or pleasurable activities become professionalized. Future research could similarly apply self-determination theory to study OnlyFans creators.
OnlyFans also offers an unusual opportunity to monetize pleasurable activities that our participants were already doing: they could quickly re-purpose their existing digital audiences or sexual content that they had initially developed for recreation (e.g., while doing hobbies like cosplay or in romantic relationships).

\subsection{Platform features to protect privacy}
As aforementioned, a vast majority of adults make and share sexual content~\cite{stasko2015sexting}. However, a great deal of literature (see e.g., \cite{hasinoff2015sexting,stasko2015sexting,de2019sexting,geeng2020usable}) focuses on the risks of creating and sharing sexual content. While Angela Jones, in her book \textit{Camming}~\cite{jones2020camming}, describes some of her webcamming participants as fulfilling their exhibitionist natures by working on platforms that offer little-to-no privacy protections, the participants in our study wanted to share in a more boundaried and discreet way. Our participants found a sense of control and autonomy over their content by using OnlyFans to share it. 

Multiple platform features contributed to the level of comfort our participants had with sharing their intimate content on OnlyFans: 
%
affordances for member verification, the ability to moderate subscribers, and the lack of searchability on both the platform and the web. Further, the paywall-based monitization model of OnlyFans, in which fans expect to have to pay to view content vs. many webcamming platforms in which streams can be viewed for free and creators receive tips much like Twitch, created an additional sense of privacy. 

\subsection{Platform features to support sexual expression}
In addition to the risks for creators in sharing their sexual content, there exist significant threats to platforms that host this content. As a result, many platforms chose to ban intimate content from being shared or eventually deplatform those who create and share such content due to pressures from funders or governments~\cite{tiidenberg2020sex}. 

OnlyFans requires its creators to identify themselves with both government ID and biometric identification. Content created with another person must be accompanied by the identification and written consent. 
Thus, OnlyFans' terms and conditions allow certain kinds of sexual content while requiring performer identification to inhibit illegal material. The performer identification to a certain extent immunizes the platform against the claim that there is illegal material on it, whilst the terms and conditions enable content that would likely get the creators deplatformed elsewhere. P21 described deplatforming thus: ``everything you worked for is just gone and then you have to rebuild and try and get in contact with customers different ways and yeah. If you don't have a website it's just horrible.'' Deplatforming is the major income risk to adult content creators. The affordances, business practices, and terms of conditions of OnlyFans therefore combine to enable less personal and structural risk of deplatforming. 

There are still limitations to OnlyFans support for sexual expression. For example, the platform now bans particular types of content (such as content shot outside) which may restrict sexual expression among some potential users~\cite{OnlyFans:Outdoors:Ban}. 
Despite these limitations, OnlyFans can serve as a valuable case study for understanding what precisely affords creators an opportunity for sexual expression and control and autonomy over that expression. Building on this work and on existing principles such as those expressed in the Sex Positive Social Media Manifesto~\cite{stardust2022manifesto}, future work should explore additional avenues for supporting sexual expression on social media. This will help us better understand how to design consensual, safe social media platforms on which people may wish to share intimate content.


\subsection{Converging and diverging needs of creators}
Our work contributes to an emerging taxonomy of creators on platforms like OnlyFans, which includes professional creators (a mix of those with existing sex-work experience and new creators) who are focused primarily on earning income through OnlyFans and more recreational creators who are leveraging OnlyFans out of a desire to engage in sexual expression. As mentioned in Section~\ref{sec:related}, Cardoso and Scarcelli also note a distinction between self-identified "amateurs" and creators using OnlyFans in a "professional capacity"~\cite{cardoso2021bodies}. Prior work finds that a majority of those already in the industry who started online sex work due to the pandemic intend to continue~\cite{hamilton2022risk}. Therefore, those formerly or simultaneously doing offline sex work also form a major cohort of OnlyFans creators. 

Prior work on webcamming also identified exhibitionism as a motivator for undertaking online  sexwork~\cite{jones2020camming}. 
%
In our work, 14 participants were motivated to join OnlyFans out of a desire to engage in sexual expression. Whilst 18 of our participants were motivated to join OnlyFans by the opportunity to monetize their content, others may have been willing to use the platform even if they did not earn income from it, because they sought a minimally stigmatized, convenient, discreet platform that permitted them to warehouse and socially share their erotic content. Continued development and refinement of this taxonomy is needed to improve the development of platforms for sexual content sharing as well as relevant policy (e.g., for protecting the rights of content creators).
%
%
%
%
%
%
%
%
%
%
%
%
\section{Conclusion}
Leveraging qualitative investigation through semi-structured interviews we examined the motivations and experiences of $22$ U.S.-based content creators who joined OnlyFans without prior experience selling sexual content. Many of the motivations of our participants to create sexual content on OnlyFans overlap with those of other professional content creators (personal interests, self-branding, and recognition) and gig workers (flexibility, autonomy, control, and management of other commitments and responsibilities), thus supporting the need to include sexual content creators as part of research and conversation on digital labor and gig work. 

Uniquely however, our participants highlight the ability for their work on OnlyFans to be enjoyable---pleasureable labor---because of the opportunity the platform provides for sexual expression as well as the harms the platform aims to prevent through privacy-related affordances (e.g., lack of searchability, paywalls) and omission of many algorithmic management features (e.g., ratings) common to other gig work platforms. 
Our results also suggest that the celebritization and corresponding increased mainstream acceptance and lessened stigma of the platform as well as income loss and changes in how leisure time was spent created by the pandemic contributed to bringing our participants---and their fans---to the platform.  We propose that OnlyFans uniquely profited off the simultaneous events of celebrity hype and the COVID-19 pandemic. It is unclear (and may be difficult to definitively determine) if the platform could have reached such saturation without either factor.

\section*{Acknowledgment}

The authors wish to thank Danielle Blunt and Maya Mundell for their contributions to recruitment and wish to thank Oshrat Ayalon, Hanna Barakat, Danielle Blunt, Kayla Booth, Angelica Goetzen, and Maya Mundell for their brainstorming and feedback on the interview protocol used in this work.

\bibliographystyle{unsrtnat}
\bibliography{arXiv-main}

\end{document}